# Comprehensive Analysis of Continuously Variable Series Reactor Using G-C Framework

Mohammadali Hayerikhiyavi, Aleksandar Dimitrovski, *Senior Member, IEEE*

*Abstract*-- Continuously Variable Series Reactor (CVSR) has the ability to regulate the reactance of an ac circuit using the magnetizing characteristic of its ferromagnetic core, shared by an ac and a dc winding to control power flow, damp oscillations and limit fault currents. In order to understand and utilize a CVSR in the power grid, it is essential to know all of its operational characteristics. The gyrator-capacitor approach has been applied to model electromagnetic coupling between the two circuits, controlled ac circuit and control dc circuit of the device. In this paper, we investigate some of the CVSR side behavior in terms of the induced voltage across the dc winding, flux density within the core's branches, and the power exchange between the two circuits during normal operation and fault conditions.

## I. Introduction

As the penetration of variable renewable energy sources continuously increases, the existing power grid is under growing stress in terms of its capability and reliability. One of the most important challenges power engineers face is preventing brownouts and blackouts due to the transmission congestions, oscillations, etc. that are result of the lack of means for comprehensive power flow control. Devices for power flow control include phase shifting transformers, switched shunt-capacitors/inductors, and various types of flexible ac transmission systems (FACTS) controllers. These devices are either costly or coarse in their functionality. Recently, saturable reactor technology based on the principle of magnetic amplifier was proposed to relieve congestion with high reliability and low costs [1-3]. Continuously Variable Series Reactor (CVSR) is a series reactor that has continuously variable reactance within its design limits. Continuous and smooth control of a large amount of power flow in the ac circuit can be achieved by controlling the bias dc current. Potentially, there can be other dynamic applications of CVSR such as oscillation damping and fault currents limitation [4, 5]. CVSR adds additional impedance into the ac circuit which can be appropriately modulated to damp oscillations as well as decrease fault currents. Hence, it is important to study the effects of the CVSR, but it is equally important to understand the impacts on it during all power system conditions.

Unlike the FACTS controllers that use high voltage and/or current components that are also part of the main power circuit [6], the dc source of the CVSR is controlled by low power-rated electronics. In [2], a relatively simple power electronics-based converter in an H-bridge composed of four IGBTs (Insulated Gate Bipolar Transistors) was described.

The paper is organized as follows: the basic concept of CVSR is briefly reviewed in Section II; the gyrator-capacitor (G-C) approach is explained in Section III; Section IV describes the simulation framework; Section V presents results from the simulations of normal operation and fault conditions to analyze the impacts on the CVSR in terms of induced voltage and power transferred into the dc windings; and conclusions are summarized in Section VI.

## II. Continuously Variable Series Reactor (CVSR)

In the saturable core reactor shown in Figure 1, an ac winding is wound on its middle leg and connected in series with an ac circuit, containing loads and voltage sources. A dc winding driven by a controllable dc source is wound on the two outer legs. The reactance of the ac winding and, hence, the ac current, is controlled by the dc current in the dc winding.

The dc current generates the bias dc flux and controls the saturation of the core, or the self-inductance of the ac winding. This inductance reaches the maximum when the core is not saturated, and the minimum when it is fully saturated (at large enough dc current). The overall reactance in the controlled ac circuit is changes by the dc current. During each half cycle of a period, the ac and dc fluxes will add in one half of the core and subtract in the other half. In other words, the magnetic permeances of the two halves can be different [7], and so will be $\frac{d\Phi_{right}}{dt}$ and $\frac{d\Phi_{left}}{dt}$ which result with the induced voltages on the right and left leg ($V_{right}, V_{leftt}$), i.e. $V_{bias} = V_{right} - V_{left}$. The frequency of this voltage is double the frequency of the voltage and current in the ac circuit.

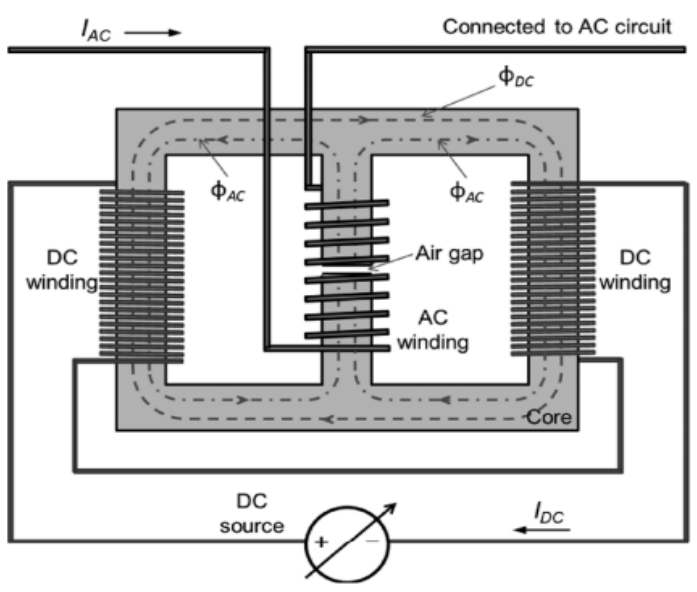

Fig. 1. Simplified schematic of CVSR [1].

## III. Gyrator-Capacitor Model

A popular approach to modelling magnetic circuits is to use electric circuit analogy [8,9]. Equivalent circuits are formed using resistors that represent reluctances and voltage sources that represent MMFs. However, there is a discrepancy between magnetic cores which store energy and resistors which dissipate energy. An energy invariant approach is offered by a gyrator-capacitor (G-C) model shown in Figure 2 [10].

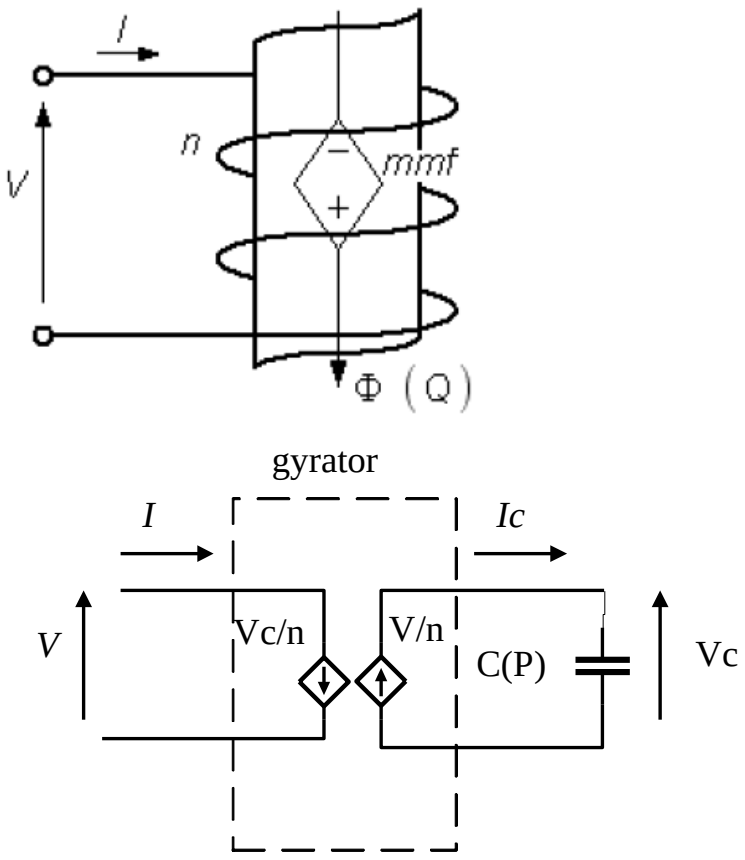

Fig. 2. magnetic circuit and its equivalent gyrator-capacitor model

From Faraday's and Ampere's laws follow (1) and (2):

$$\frac{d\Phi}{dt} = -\frac{V}{n} \quad (1)$$

$$I = \frac{mmf}{n} \quad (2)$$

In the gyrator capacitor (G-C) approach, the analogy between MMF and voltage, and current and rate-of-change of magnetic flux $\frac{d\Phi}{dt}$ are described by (3) and (4):

$$V_c \equiv mmf \quad (3)$$

$$I_c = \frac{d\Phi}{dt} \quad (4)$$

The model of a simple magnetic device using the G-C method is also shown in Fig. 2. Nonlinear permeances representing nonlinear magnetic paths are modeled as capacitors and windings are represented by gyrators. The permeances are calculated from the material B-H curve and the geometric parameters as given by (5):

$$\rho = \frac{\mu_r \mu_0 l}{A} \quad (5)$$

where $\mu_0 = 4\pi\times10^{-7}$, $\mu_r$ is the magnetic permeability of the material, $A$ is the cross-sectional area of the path, and $l$ is the mean length of path.

## IV. Simulation

The G-C equivalent model of the CVSR in Fig. 1 is shown in Fig. 3. The three ferromagnetic legs are modeled with three nonlinear permeances, and a single linear permeance models the air gap in the middle leg as shown in Fig. 3. The dc controlled circuit is connected to an ideal dc source and the two gyrators in this circuit represent each coil of the dc winding. The ac circuit consists of a load and a voltage source with the gyrator in the middle representing the AC winding [11,12].

The model was created and simulated using SIMULINK® and MATLAB® [13]. The device and circuit parameters are summarized in Table I. The ferromagnetic core material assumed is M36 with specifications taken from [14].

The fringing flux due to the air gap [15, 16] has been taken into account in the air gap permeance.

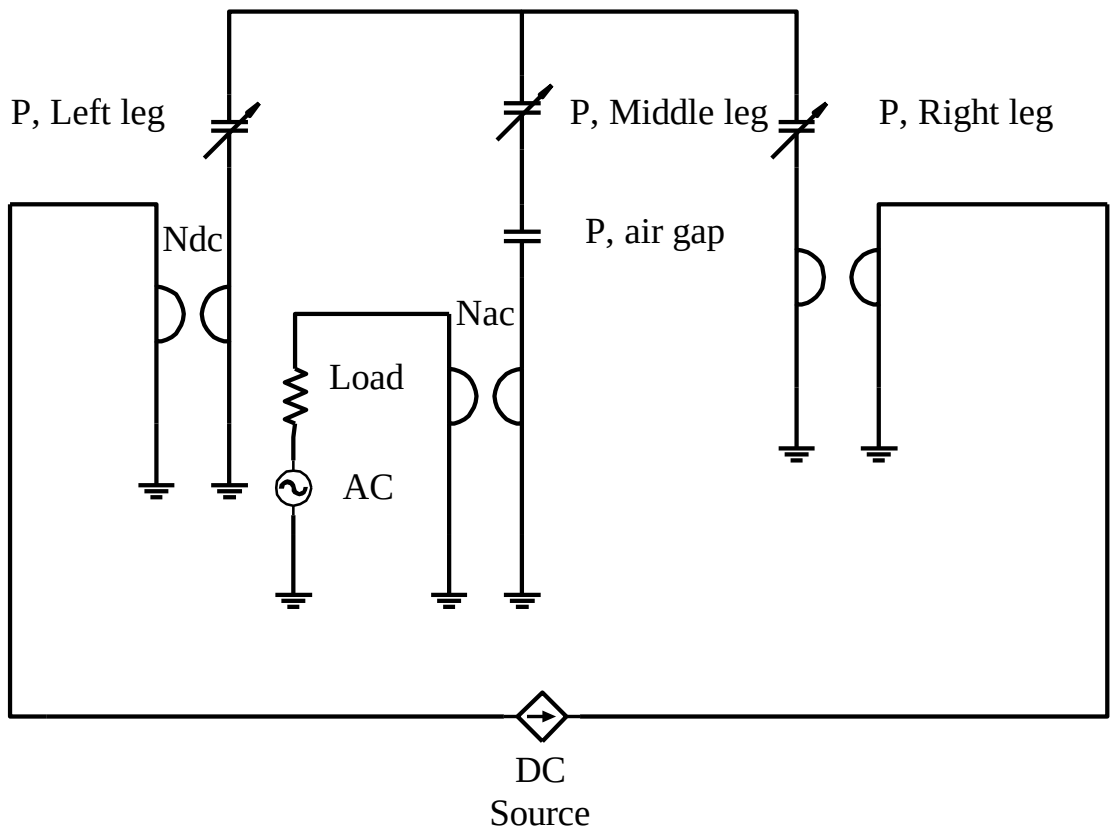

Fig. 3. Gyrator-Capacitor Model in SIMULINK

TABLE I MODEL PARAMETERS

| Parameter | Description | Value |
|---|---|---|
| $l_m$ | Mean Length of middle leg | 45.72cm |
| $l_{out}$ | Mean Length of outer leg | 86.36cm |
| $H_{ag}$ | Height of air gap | 0.2014cm |
| $A$ | Cross-Section Area | $0.0103m^2$ |
| $N_{dc}$ | Number of DC turns of dc winding | 450 |
| $N_{ac}$ | Number of AC turns of ac winding | 300 |
| $V$ | Voltage source | 2.4Kv (RMS) |
| $R$ | Load Resistance | 100Ω |
| $L$ | Load Inductance | 130 mH |
| $P.F$ | Power Factor | 0.9 |
| S | Device rated power | 50KVA |
| $B_{sat}$ | Saturation point | 1.34 Tesla |

## V. Results

### A. Nominal conditions

In this case, four different dc bias currents are applied: 0A, 20mA, 75mA (critical current) , and 5 A. For the critical current, the operating point is at the knee of the B-H curve, one of the outer legs is in saturation and the other one is in the unsaturated region. Several ac currents are chosen to model all operation conditions in the G-C model.

The waveforms in Figures. 4 and 5 represent the terminal voltage and core magnetic inductions (flux densities) for the G-C model with dc bias of 0 A. The terminal voltage is purely sinusoidal. The inductions through the outer legs are equal and in phase. Thus, the induced voltage across dc windings will be zero.

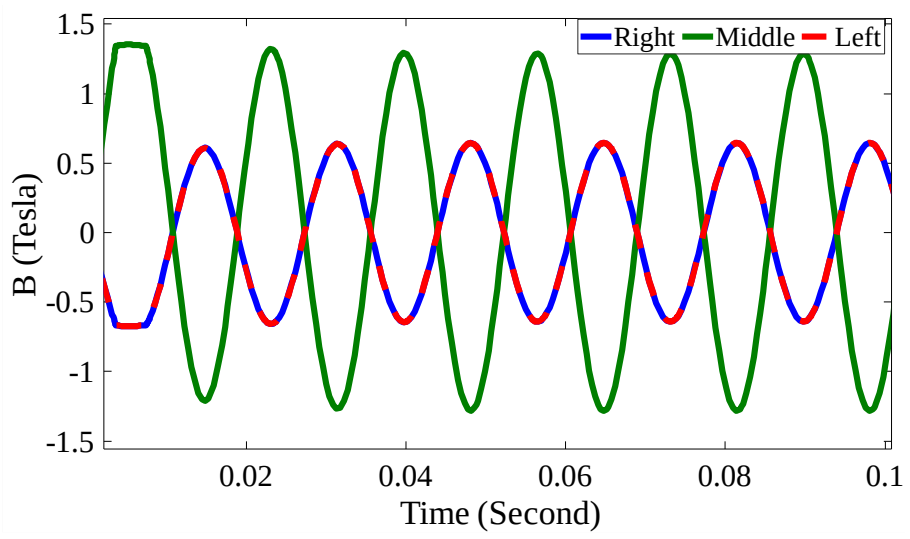

Fig. 4. Inductions through all legs (DC=0)

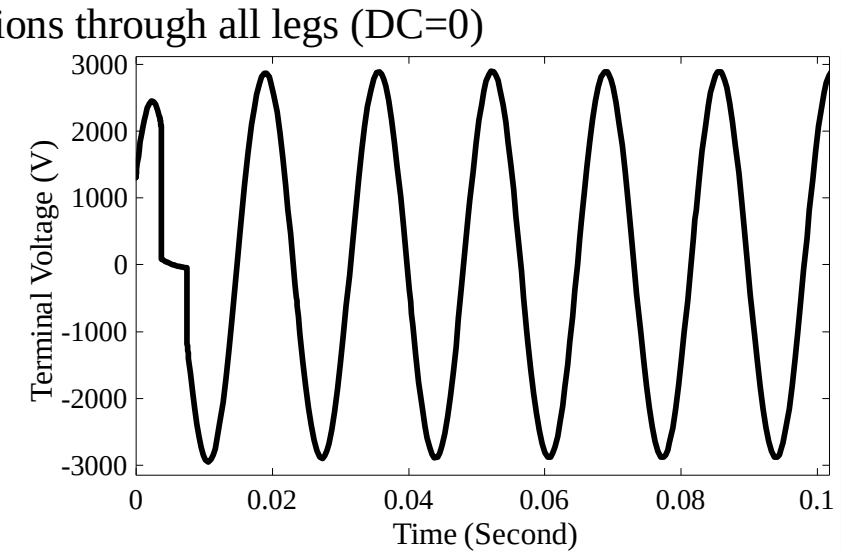

Fig. 5. Induced voltage across AC winding (DC=0A)

The current through the ac winding is shown in Figure 6. It can also be easily calculated from the voltage source and the equivalent circuit impedance which includes the load and the CVSR reactance in series.

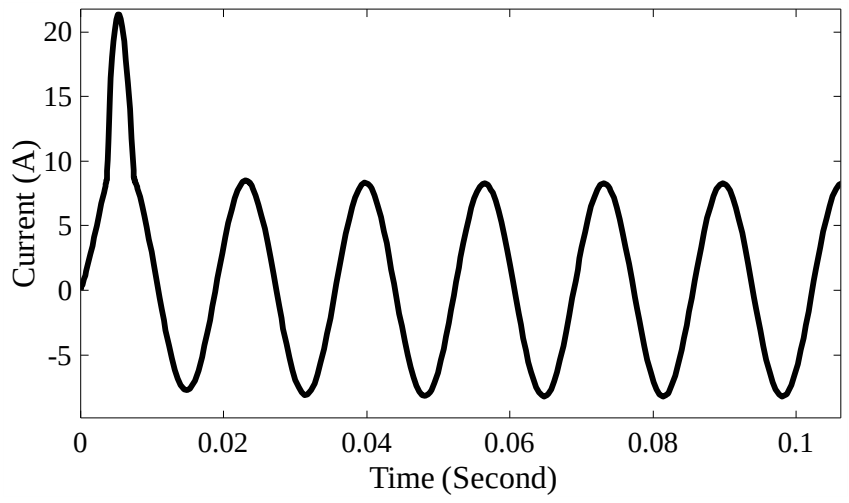

Fig. 6. Current in AC winding (DC=0A)

When dc bias is zero, there is no power transferred into the dc windings, because the AC flux will be divided to the outer legs equally. The induced voltage is equivalent to the difference between the flux rates of the outer legs:

$$E_{emf} = N_{dc}.\left(\frac{d\Phi_{right}}{dt} - \frac{d\Phi_{left}}{dt}\right) \quad (7)$$

$E_{emf}$: induced voltage

$N_{dc}$: Number of DC turns of dc winding

$\frac{d\Phi}{dt}$: Flux rate

Therefore, the induced voltage across DC winding will be zero.

In Figure7, the magnetic inductions have a dc component resulting from the 20mA dc bias current. The inductions through the outer legs are no longer equal, but they are still in phase. So, the induced voltage across the dc winding will be zero. Thus, the power transferred to dc winding is also zero as shown in Figure 8.

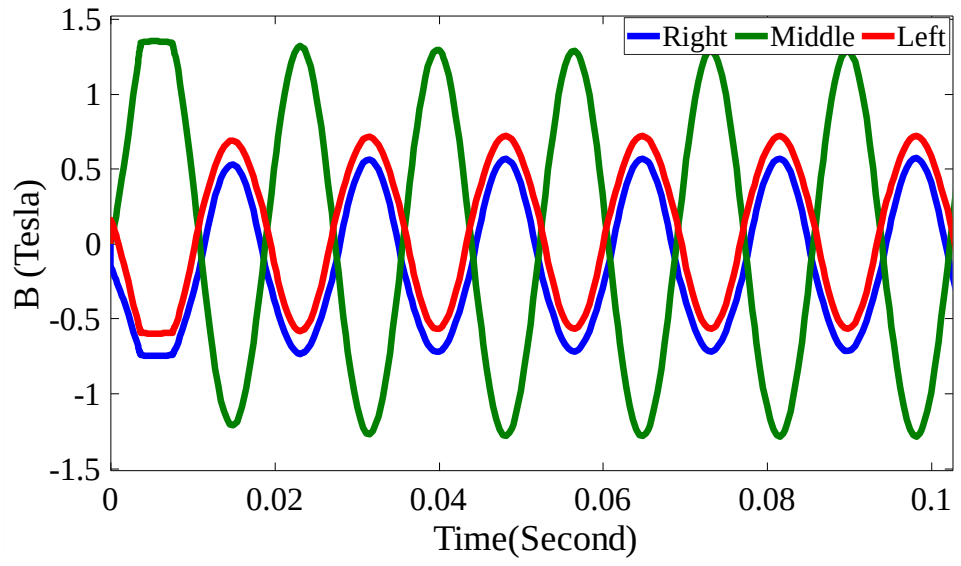

Fig. 7. Inductions through all legs (DC=20mA)

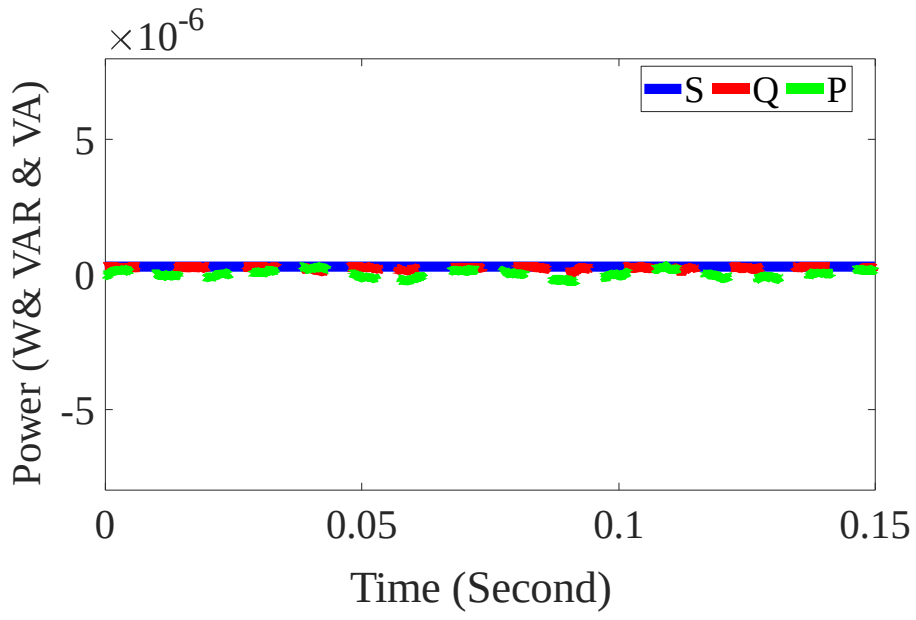

Fig. 8. Powers Transferred to the DC winding (DC=20mA)

In Figure 9, the dc bias is increased to 75 mA, causing enough dc offset in the fluxes in the outer legs. The outer leg inductions ($B$) are distorted as the core enters into saturation. Therefore, the outer inductions are no longer equal in terms of shape and phase. At any time, the flux produced by the middle leg will increase the permeance of one of the outer legs and reduce the permeance of the other one. So, the rate of change of the fluxes passing through the outer legs will differ and the induced voltage on the outer windings will not be zero. As can be seen from Figure10, the frequency of this induced voltage across dc windings is double the frequency of the voltage source.

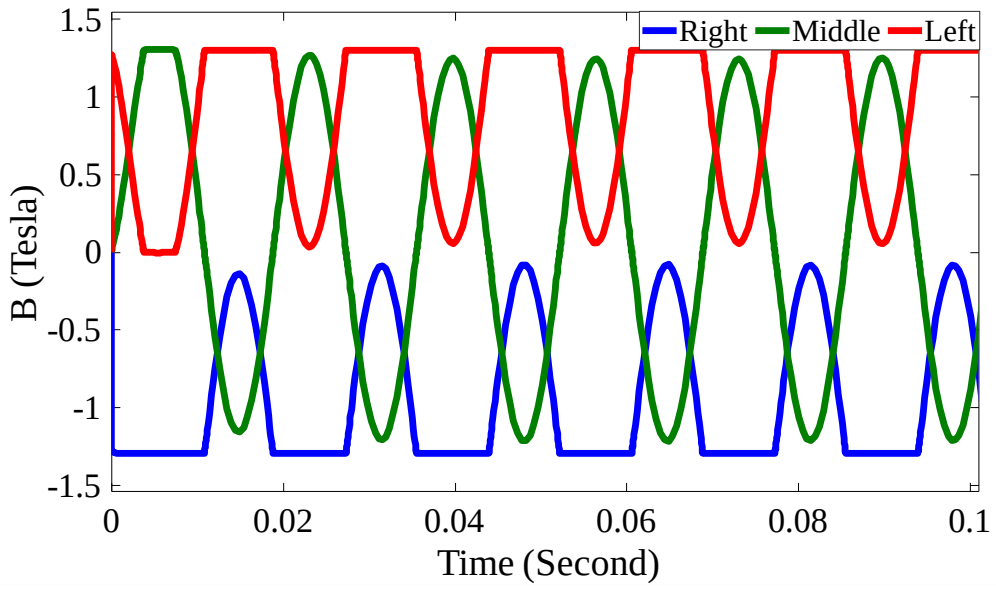

Fig. 9. Inductions through all legs (DC=75mA)

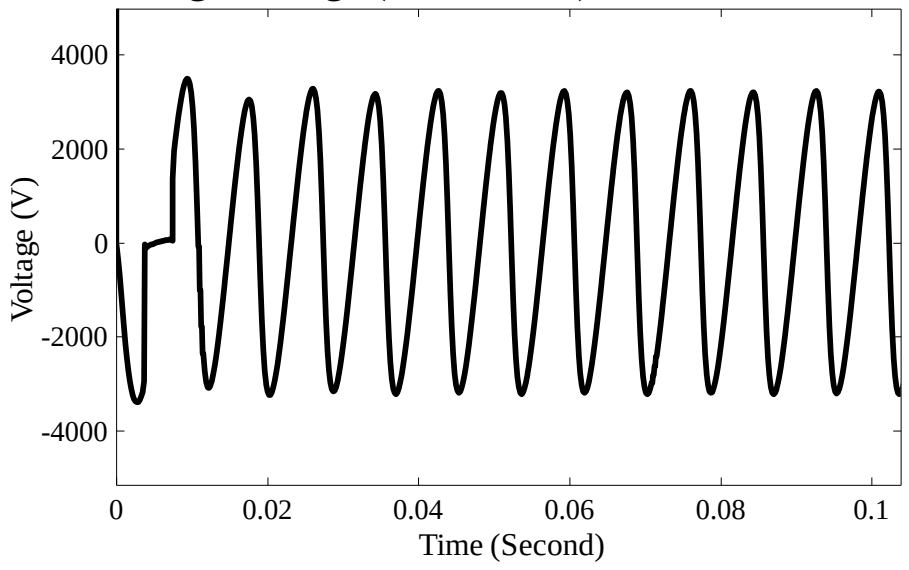

Fig. 10. Induced voltage across DC windings (DC=75mA)

The current in the ac winding stays the same when dc current is 0A, 20mA, or 75mA. These currents are not high enough to drive the outer legs into saturation and, overall, the ac circuit virtually stays the same.

In Figure 11, all the powers (real, reactive, and apparent) transferred into the dc control circuit are shown.

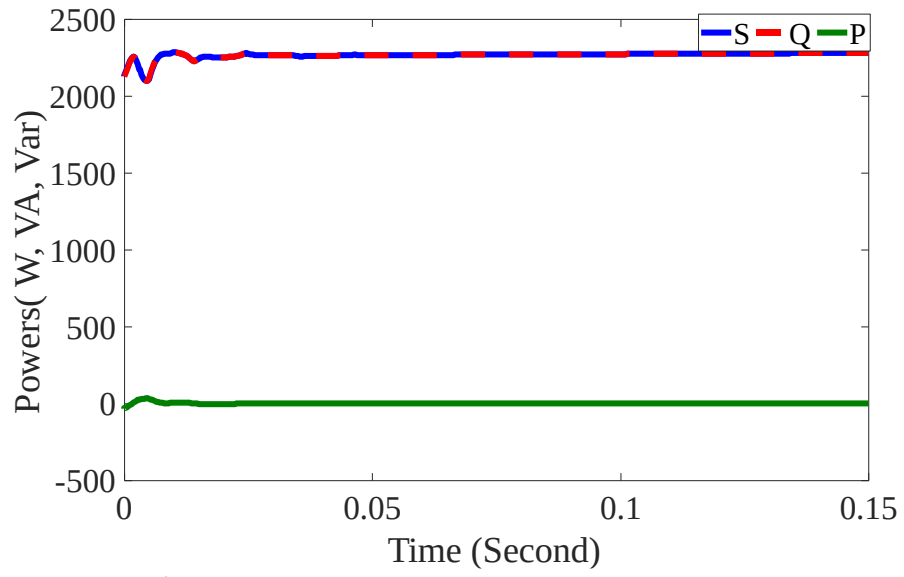

Fig. 11. Powers Transferred to the DC winding (DC=75mA)

In Figure 12, the CVSR completely goes into saturation, because of high DC offset. The ac induction (*B*) in the center decreased due to the high reluctance of the outer legs. but stays purely sinusoidal.

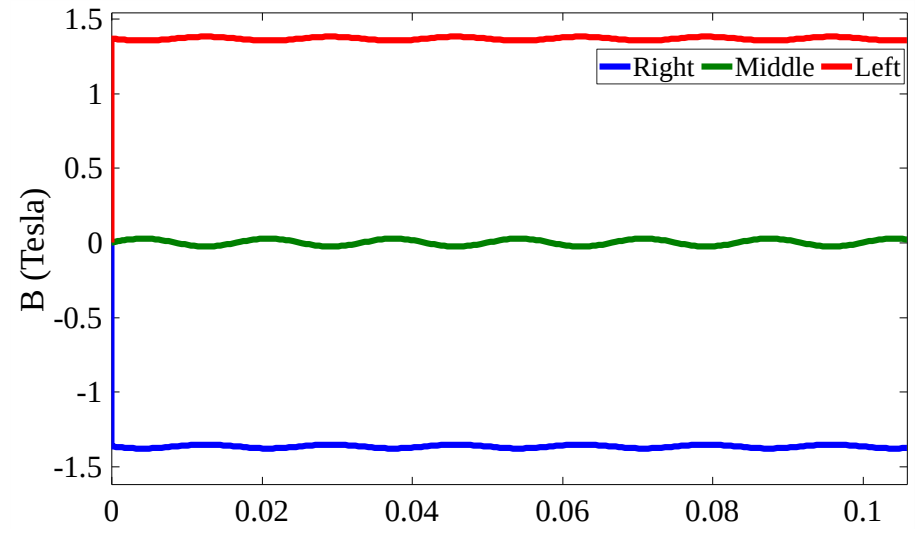

Fig. 12. Inductions through all legs (DC=5A)

Current through the ac winding is shown in Figure 13. Due to the complete saturation of the core, reactance in the ac circuit will be negligible, so these determined by the load impedance.

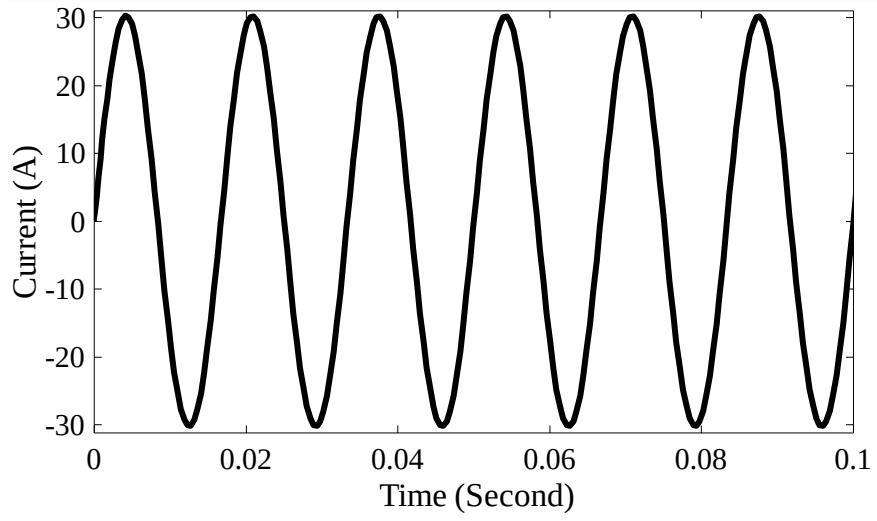

Fig. 13. Current in AC winding (DC=5A)

Since, the outer legs are fully saturated, the induced voltage on the left and right leg are equal and, the induced voltage across the dc winding will be zero. Thus, the power transferred into the dc side is also zero.

## B. Fault Conditions

In this part, initially, the system works under normal condition for 2 cycles when a fault is initiated. It has been assumed that the fault occurs at 90 percent of the total load impedance in the ac circuit. Again, four different dc biases are applied: 0, 20mA, 75mA, and 5 A.

The waveforms in Figures 14, 15, and 16 represent the terminal voltage, core inductions and current in ac winding, respectively, for the CVSR with dc bias of 0 A. The terminal voltage is purely sinusoidal. The induced voltage across dc winding will be zero, because the ac flux will be divided between the outer legs equally. Thus, the power transferred into the dc circuit is zero. In Figure 15, the flat regions in the induction through the middle leg are due to saturation. Accordingly, the terminal voltage and current are distorted at the corresponding time instances.

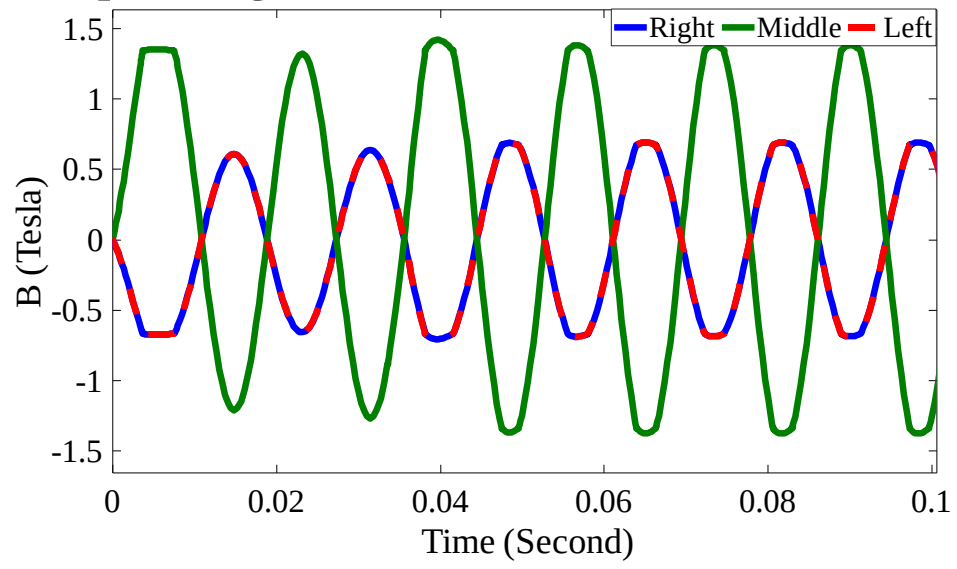

Fig. 14. Inductions through all legs (DC=0)

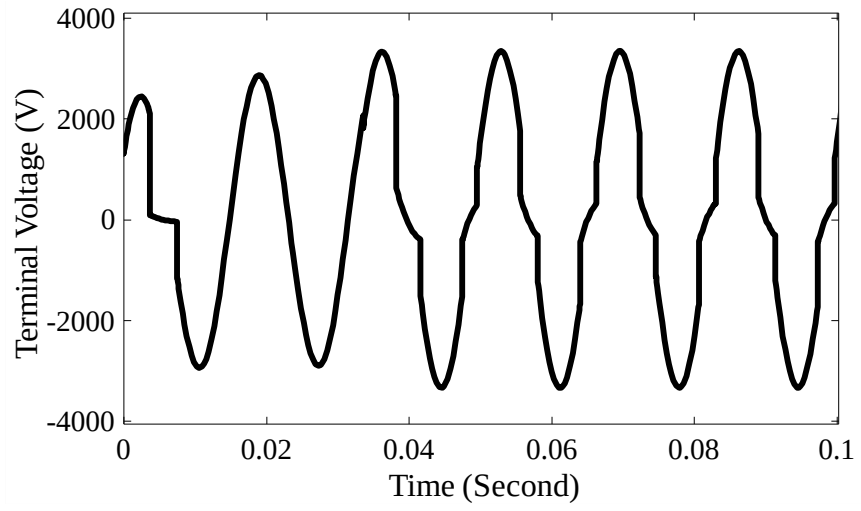

Fig. 15. Induced voltage across AC winding (DC=0A)

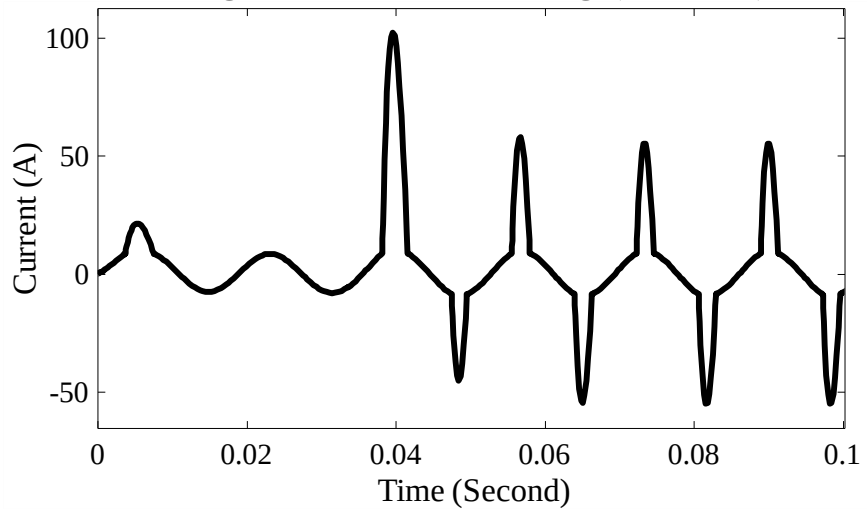

Fig. 16. Current in AC winding (DC=0A)

In Figure 17, the inductions through all the legs are shown. Similarly to the previous case (normal condition with DC=20 mA), since the inductions through the outer legs are in phase in a linear region, the induced voltage across dc winding will be zero.

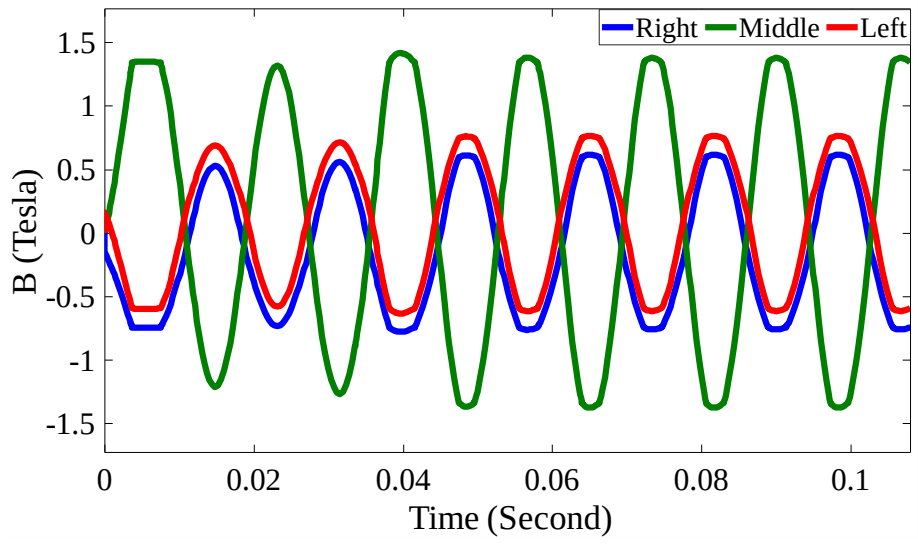

Fig. 17. Inductions through all legs (DC=20mA)

In Figure 18, the DC is 75 mA. The middle leg is unsaturated, but when the fault occurs after 2 cycles, it enters into saturation. The flat regions in the induction (B) through the middle leg are due to the saturation. In Figure 19, it can be observed the frequency of the induced voltage across dc windings is double the frequency of input voltage. Rolling window is able to capture the transient behavior of operation conditions accurately. Therefore, this approach is implemented to determine the transferred power to the DC windings (see Figure 20). Again, the current in the ac windings is identical when dc is equal to 0A, 20mA, and 75 mA.

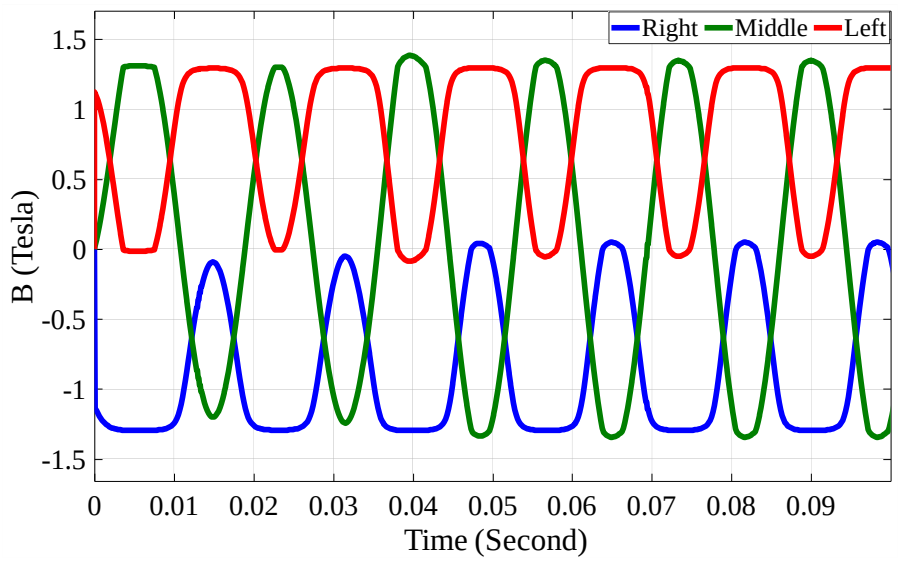

Fig. 18. Inductions through all legs (DC=75mA)

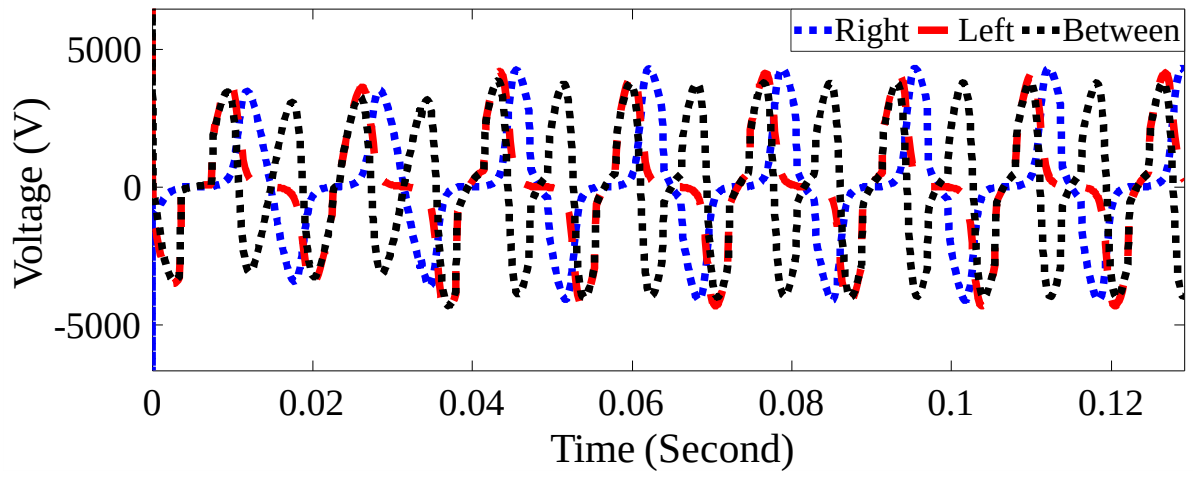

Fig. 19. Induced voltage on DC windings (DC=75mA)

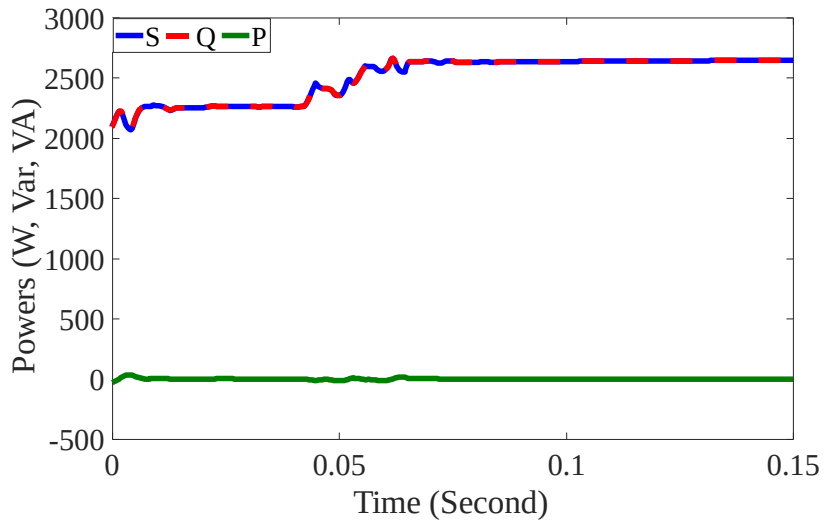

Fig. 20. Powers Transferred to the DC winding (DC=75mA)

In the pre-fault period, the dc bias is 5A, and the outer legs are completely saturated. When the fault occurs after two cycles, the high ac current leads outer legs out of fully saturation (see Figure 21). Additionally, the induced voltage across dc windings and the power transferred to the dc winding are shown in Figure22 and 23, respectively. The asymmetry observed in the induced voltage across dc windings (Figure22) is the result of saturation of the middle leg.

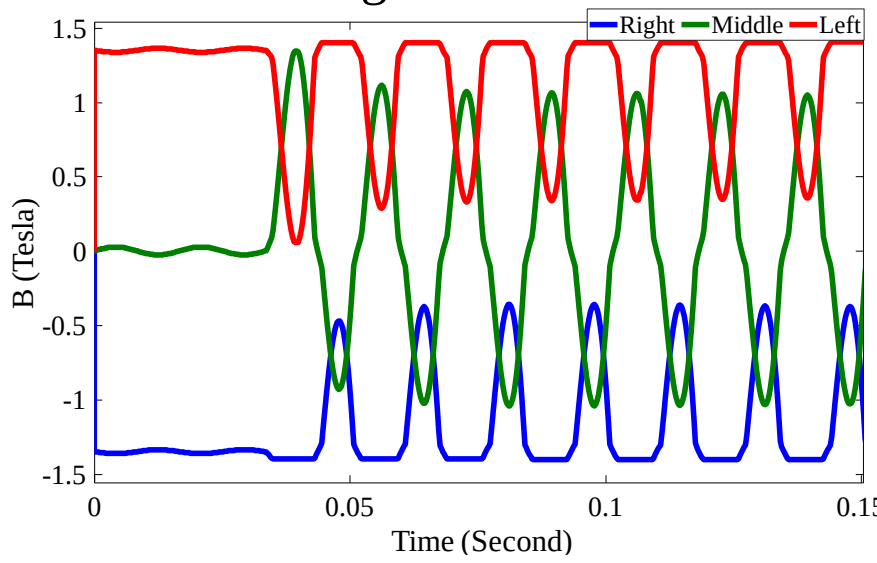

Fig. 21. Inductions through all legs (DC=5A)

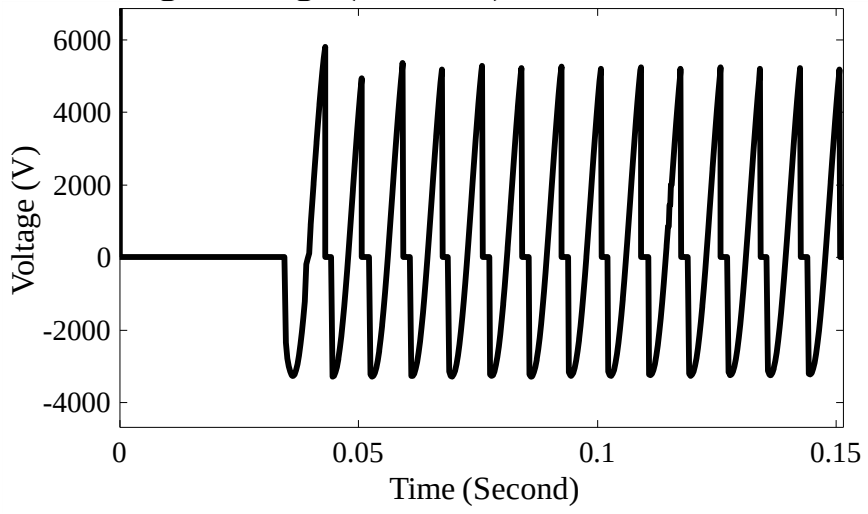

Fig. 22. Induced voltage across DC windings (DC=5A)

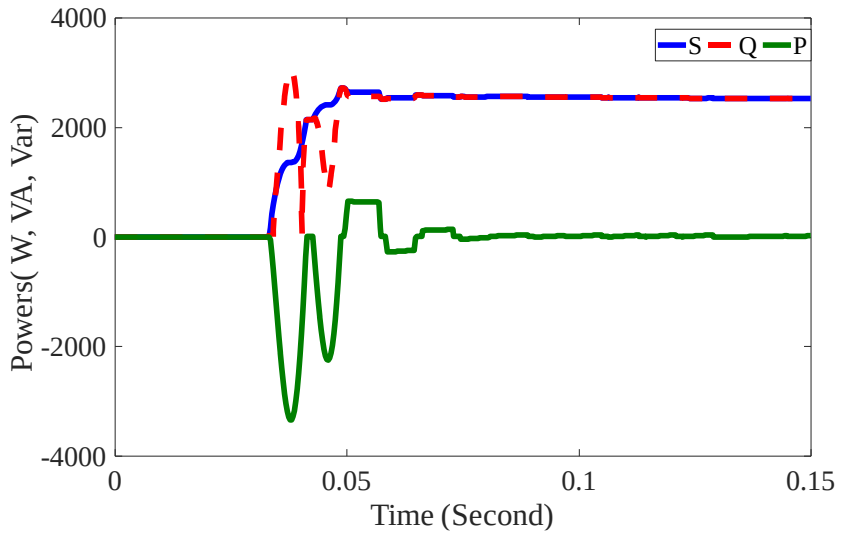

Fig. 23. Powers Transferred to the DC winding (DC=5A)

## VI. Conclusion

The present study is devoted to simulate the impact on the Continuously Variable Series Reactor (CVSR), in particular the induced voltage across the dc winding, inductions through all the legs, and the power transferred into the dc circuit.

Gyrators/capactiros provide an analogy between the magnetic circuits and the electric circuits. Nonlinear capacitors, or permeances, are used to model the nonlinear core characteristics. Simulations of a G-C model is based on the configuration of the three legged CVSR under nominal and fault conditions for different value of DC current.

Future work will investigate the possible operation of this device as a fault current limiter.